\documentclass[twocolumn,preprintnumbers,amsmath,amssymb]{revtex4}
\usepackage{array}
\usepackage{booktabs}
\usepackage{tabu}
\usepackage{dcolumn}
\usepackage{amsmath}
\usepackage{amsfonts}
\usepackage{amssymb}
\usepackage{graphicx,color}
\usepackage[colorlinks={true}]{hyperref}
\usepackage{graphicx}
\usepackage{subfigure}
\usepackage{graphicx}% Include figure files
\usepackage{dcolumn}% Align table columns on decimal point
\usepackage{bm}% bold math

\def\be{\begin{equation}}
  \def\ee{\end{equation}}
\def\bea{\begin{eqnarray}}
\def\eea{\end{eqnarray}}
\def\f{\frac}
\def\n{\nonumber}
\def\l{\label}
\def\p{\phi}
\def\o{\over}
\def\R{\rho}
\def\pa{\partial}
\def\om{\omega}
\def\na{\nabla}
\def\P{\Phi}
\begin{document}

\title{The entropy production of thermal operations}% Force line breaks with \\

\author{H. Dolatkhah}
\author{S. Salimi}
\email{shsalimi@uok.ac.ir}
\author{A. S. Khorashad}
\affiliation{
Department of Physics, University of Kurdistan, P.O.Box 66177-15175, Sanandaj, Iran}
\author{S. Haseli}
\affiliation{Department of Physics, Urmia University of Technology, Urmia, Iran.\\}
\date{\today}% It is always \today, today,

\def\be{\begin{equation}}
  \def\ee{\end{equation}}
\def\bea{\begin{eqnarray}}
\def\eea{\end{eqnarray}}
\def\f{\frac}
\def\n{\nonumber}
\def\l{\label}
\def\p{\phi}
\def\o{\over}
\def\R{\rho}
\def\pa{\partial}
\def\om{\omega}
\def\na{\nabla}
\def\P{\Phi}
%\nofiles

%=============================================================%
%=============================================================%
%============== Abstract =======================================%
%=============================================================%
%=============================================================%
\begin{abstract}
According to the first and second laws of thermodynamics and the definitions of work and heat, microscopic expressions for the non-equilibrium entropy production have been achieved. Recently, a redefinition of heat has been presented in [\href{Nature Communicationsvolume 8, Article number: 2180 (2017)}{Nat. Commun. 8, 2180 (2017)}]. We are going to determine how this redefinition of heat could affect the expression of  the entropy production. Utilizing this new definition of heat, it could be found out that there is a new expression for the entropy production for thermal operations. It could be derived if the initial state of the system and the bath is factorized, and if the total entropy of composite system is preserved, then the new entropy production will be equal to mutual information between the system and the bath. It is shown that if the initial state of the system is diagonal in energy bases, then the thermal operations cannot create a quantum correlation between the system and the bath.

\end{abstract}
%\pacs{04.50}
%\keywords{keyword.}%Use showkeys class option if keyword
\maketitle

%%%%%%%%%%%%%%%%%%%%%%%%%%%%%%%%%%%%%%%%%%%%%%%%%%%%%%%%%%%%%%%%%%%%%%%%%%%%
%%%%%%%%%%%%%%%%%%%%%%%%%%%%%%%%%%%%%%%%%%%%%%%%%%%%%%%%%%%%%%%%%%%%%%%%%%%%
%%%%%%%%%%%%%%%%%%%%%%%%%%%%%%%%%%%%%%%%%%%%%%%%%%%%%%%%%%%%%%%%%%%%%%%%%%%%
%%%%%%%%%%%%%%%%%%%%%%%%%%%%%%%%%%%%%%%%%%%%%%%%%%%%%%%%%%%%%%%%%%%%%%%%%%%%
%============  Sec.I (Introduction)  =======================================
%%%%%%%%%%%%%%%%%%%%%%%%%%%%%%%%%%%%%%%%%%%%%%%%%%%%%%%%%%%%%%%%%%%%%%%%%%%%
%%%%%%%%%%%%%%%%%%%%%%%%%%%%%%%%%%%%%%%%%%%%%%%%%%%%%%%%%%%%%%%%%%%%%%%%%%%%
%%%%%%%%%%%%%%%%%%%%%%%%%%%%%%%%%%%%%%%%%%%%%%%%%%%%%%%%%%%%%%%%%%%%%%%%%%%%
%%%%%%%%%%%%%%%%%%%%%%%%%%%%%%%%%%%%%%%%%%%%%%%%%%%%%%%%%%%%%%%%%%%%%%%%%%%%
\section{INTRODUCTION}	%) A SECTION HEADING
The conservation of energy in thermodynamic systems is the topic of the first law of the thermodynamics, which states that every increase in the internal energy of a system is produced in one of two ways (or both): (a) by means of work performed on the system and/or (b) by means of heat absorbed by the system. Heat is defined as the flow of energy from the environment, normally considered as a thermal bath at a certain temperature, to a system, in some way different from work. Irreversible processes are described by the second law of thermodynamics, the non-negativity of  the entropy production $\Delta S^{irr}\geq0$ is called the Clausius inequality that is one of the various ways of expressing the second law. According to the second law of thermodynamics, the entropy production is always non-negative, being zero only when the system and the environment are in thermal equilibrium. Recently, thermodynamic behavior within quantum mechanical systems has received a huge interest. Therefore, having a clear understanding about the fundamental concepts such as work, heat and the second law in the quantum domain has been the main topic of many research in quantum thermodynamics \cite{Gelbwaser,Weimer,Salmilehto,Hossein-Nejad,Gallego,Jarzynski,Jarzynski2,Borhan,Borhan1}. \\
Some microscopic expressions for the non-equilibrium entropy production have been derived in open and closed quantum systems \cite{Deffner,Deffner2,Esposito,Esposito2,chen}.

In order to understand the foundations of thermodynamics in quantum domain, thermodynamics is viewed as a resource theory. There are different models for the resource theories of thermodynamics \cite{Gour,Goold,Nelly}, which vary mostly on the set of allowed operations. One of the most important models is the resource theory of thermal operations \cite{Brandao,Brandao2,Horodecki,Janzing}.

Recently, a redefinition of heat has been presented \cite{Bera} where the authors introduced a redefine of heat by properly accounting for the information flow and thereby restoring Landauer's erasure principle. During the presented work, it aims to determine how this redefinition of heat could affect the expression of the entropy production for thermal operations. Since, microscopic expressions for the non-equilibrium entropy production is based on the definition of heat and work, it is expected that any new definition of them might dramatically change the expression of the entropy production. \\

The paper is organized as follows: In Sec. \ref{Sec2} a brief review on thermal operations and the heat definitions are presented. Expressions for the non-equilibrium entropy production are obtained by using the heat definitions, in Sec. \ref{Sec3}. The role of  quantum coherence in the entropy production for thermal operations is studied in Sec. \ref{Sec4}. The final section, Sec. \ref{conclusion}, is devoted to the conclusions and discussions.

\section{Preliminary}\label{Sec2}
%\section{thermal operations}\label{Sec2}
 Thermal operations (TO) $\lbrace\varepsilon_{T}\rbrace$ consist of all maps on a system with Hamiltonian $H_{S}$ and a state $\rho_{S}$ that can be written as \cite{Brandao,Horodecki,Janzing}:
\begin{equation}\label{definationTO}
\rho'_S=\varepsilon_{T}(\rho_{S})=Tr_{B}(\rho^{\prime}_{SB}),\quad \rho^{\prime}_{SB}=U\rho_{S}\otimes\gamma_{B}U^{\dagger},
\end{equation}
where
\begin{enumerate}
\item $U$ is an energy-preserving unitary satisfying
\begin{equation}\label{unitary}
[U,H_{S}+H_{B}]=0,
\end{equation}
\item $\gamma_{B}$ is a thermal state of the bath at some fixed  inverse temperature:
\begin{equation}\label{thermal state}
\gamma_{B}=\frac{e^{-\beta H_{B}}}{Z_{B}},
\end{equation}
where $\beta$ is a fixed inverse temperature of the bath (throughout the paper we assume $k_{B}=1$), $Z_{B}=tr(e^{-\beta H_{B}})$ is known as the partition function, $H_{B}$ is an
arbitrary bath Hamiltonian.
\end{enumerate}
There are two important properties for TO which are \cite{Rudolph,Rudolph2}:
\begin{enumerate}
\item  $\lbrace\varepsilon_{T}\rbrace$ are time-translation symetric \citep{marvian}; i.e.,
\begin{equation}\label{PTO}
\varepsilon(e^{-iH_{s}t}\rho e^{iH_{s}t})=e^{-iH_{s}t}\varepsilon(\rho)e^{iH_{s}t},
\end{equation}
\item the thermal state is preserved by $\lbrace\varepsilon_{T}\rbrace$,
\begin{equation}\label{PTO2}
\varepsilon_{T}(\gamma_{B})=\gamma_{B}.
\end{equation}
\end{enumerate}
%%%%\section{Definiton of heat}\label{Sec3}
The non-equilibrium free energy for a system in a state $\rho_{S}$, with Hamiltonian $H_{S}$, and with respect to a thermal bath at temperature $T$ is defined as:
\begin{equation}\label{deffree energy}
F(\rho_{S})=E_{S}-TS(\rho_{S}),
\end{equation}
where $E_{S}=tr(H_{S}\rho_{S})$ is internal energy, $S(\rho_{S})=-tr\big( \rho_{S}\ln(\rho_{S}) \big)$
is the von Neumann entropy. If we use $ H_{S}=-\frac{1}{\beta}\big(\ln( \rho^{eq}_{S}) + \ln( Z_{s}) \big)$, the non-equilibrium free energy can be written
\begin{equation}\label{deffree energy2}
F(\rho_{S})=F_{eq}+TS(\rho_{S}\Vert\rho^{eq}_{S}),
\end{equation}
where $F_{eq}=\frac{-1}{\beta} \ln(Z)$ is free energy at thermal equilibrium, and $S(\rho_{S}\Vert\rho^{eq}_{S})$ is the relative entropy. Usually, the change in the internal energy of the bath is defined as heat \cite{Reeb,Jennings}, i.e.,
\begin{equation}\label{defheat}
\Delta \bar{Q}= -\Delta E_{B}.
\end{equation}
However, recently a new definition of heat is introduced \cite{Bera}, where they state that heat should be considered as information flow. Assume there is a thermal bath with Hamiltonian $H_{B}$ at temperature $T$ that is exhibited by the thermal state $\gamma_{B}$. For a process that transforms the thermal bath $ \rho_{B}\longrightarrow\rho^{\prime}_{B}$ with the fixed Hamiltonian $H_{B}$ the heat is quantified as \cite{Bera}:
\begin{equation}\label{defheat2}
\Delta Q=\frac{-1}{\beta} \Delta S_{B},
\end{equation}
where $\Delta S_{B}=S(\rho^{\prime}_{B})-S(\rho_{B})$ is the change in von Neumann entropy of bath. Comparing Eq.\;(\ref{defheat}) and Eq.\;(\ref{defheat2}), we have
\begin{equation}\label{aa}
\Delta Q=\Delta \bar{Q}+\Delta F_{B}.
\end{equation}
Note that, if the bath deviates from thermal equilibrium by small variation during TO, i.e., $\rho^{\prime}_{B}\sim\rho^{eq}_{B}+\epsilon$, then $\Delta F_{B}=TS(\rho^{\prime}_{B}\Vert\rho^{eq}_{B})\sim\epsilon^{2}$ which in the limit of large baths $\Delta F_{B}\longrightarrow 0$. Therefore, both definitions coincide.

%%%%%%%%%%%%%%%%%%%%%%%%%%%%%%%%%%%%%%%%%%%%%%%%%
%%%%%%%%%%%%%%%%%%%%%%%%%%%%%%%%%%%%%%%%%%%%%%%%%
%%%%%%%%%%%%%%%%%%%%%%%%%%%%%%%%%%%%%%%%%%%%%%%%%
%%%%%%%%%%%%%%%%%%%%%%%%%%%%%%%%%%%%%%%%%%%%%%%%%
%%%%%%%%%%%%%%%%%%%%%%%%%%%%%%%%%%%%%%%%%%%%%%%%%
%%%%%%%%%%%%%%%%%%%%%%%%%%%%%%%%%%%%%%%%%%%%%%%%%
\section{The effect of redefining heat on the entropy production for thermal operations}\label{Sec3}
The change in entropy of the system $\Delta S_{s}$ includes a reversible and a irreversible contributions when a system experiences a dynamical process. The reversible contribution is due to heat flow which is addressed as the entropy flow $\Delta S^{rev}=\beta\Delta Q$, and the irreversible contribution is called entropy production $\Delta S^{irr}$:
\begin{equation}\label{entropyproduction}
\Delta S_{s}=\Delta S^{irr}+\Delta S^{rev}.
\end{equation}
In the following lines, the entropy production for the thermal operations is extracted by using the definitions of the heat presented in the previous section. Based on the usual definition of heat Eq.\;(\ref{defheat}) and using Eq.\;(\ref{entropyproduction}), the entropy production for the TO will be:
\begin{equation}\label{epe}
\bar{\Delta S^{irr}}=-\beta\Delta F_{S}=S(\rho_{S}\Vert\rho^{eq}_{S})-S(\rho^{\prime}_{S}\Vert\rho^{eq}_{S}),
\end{equation}
which is the familiar form of the entropy production for TO \cite{Deffner,Santos} (see Appendix for more details). It is clear that with respect to the contraction of relative entropy \cite{Nielsen}, $\bar{\Delta S^{irr}} $ is always non-negative. As a result
\begin{equation}\label{freeupper}
\Delta F_{S} \leq 0,
\end{equation}
which means that under TO, free energy is decreasing \cite{Rudolph,Brandao2}.
According to the definition of TO, Eq.\;(\ref{epe}) can be written as \cite{Esposito,Esposito2,chen}:
\begin{equation}\label{epem}
\bar{\Delta S^{irr}}=S(\rho^{\prime}_{B}\Vert\rho^{eq}_{B})+I(\rho^{\prime}_{SB}),
\end{equation}
where $I=S(\rho^{\prime}_{S})-S(\rho^{\prime}_{B})-S(\rho^{\prime}_{SB})$ is mutual information between the system and the bath. According to this equation the mutual information and the change in the state of the bath contribute to the entropy production.

We now want to see how the relation of the entropy production will be affected by changing the definition of heat. If we use the Eq.\;(\ref{defheat2}) to define the heat, there is
\begin{eqnarray}\label{epe2}
\Delta S^{irr}&=&-\beta(\Delta F_{S}+\Delta F_{B})\\
&&=S(\rho_{S}\Vert\rho^{eq}_{S})-S(\rho^{\prime}_{S}\Vert\rho^{eq}_{S})-S(\rho^{\prime}_{B}\Vert\rho^{eq}_{B}),\nonumber
\end{eqnarray}
more details could be found in Appendix. Applying the definition of TO, Eq.\;(\ref{epe2}) can be written as follows:
\begin{equation}\label{epem2}
\Delta S^{irr}=I(\rho^{\prime}_{SB}),
\end{equation}
and since $I$ is always non-negative, hence $\Delta S^{irr}\geq0$ as well.
In contrast to the Eq.\;(\ref{epem}) where mutual information and change in the state of bath participate in the entropy production, here only mutual information plays a role in the entropy production. On the other hand, from Eq.\;(\ref{epe2}) an upper bound for $\Delta F_{S}$ is concluded
\begin{equation}\label{freeupper2}
\Delta F_{S}\leqslant-\Delta F_{B}.
\end{equation}
Since $-\Delta F_{B} \leqslant 0$, the above relation is tighter than Eq.\;(\ref{freeupper}).
It is important to note that for entropy preserving operations \cite{Bera} with this condition that initial state of the composite system SB is factorized, the relation (\ref{epem2}) is also true; refer to Appendix.

\section{Quantum coherence and the entropy production}\label{Sec4}
Recently, it has been shown that the non-equilibrium free energy can be written as follows \citep{Plastina,Santos}:
\begin{equation}\label{freedivided}
F(\rho_{S})=F_{eq}+TS(p_{S}\Vert p^{eq}_{S})+TC(\rho_{S}),
\end{equation}
where $S(p_{S}\Vert p^{eq}_{S})=\sum_{i}p_{E_{i}}\ln p_{E_{i}}/p^{eq}_{E_{i}}$ is the Kullback-Leiber divergence of the classical probability distribution entailed by the populations $ p_{S}=\lbrace p_{E_{i}}\rbrace$ from that of the equilibrium state $p^{eq}_{S}=\lbrace p^{eq}_{E_{i}}\rbrace$, $p_{E_{i}}=\langle E_{i}\vert \rho_{S}\vert E_{i}\rangle$. Moreover, $C(\rho_{S})$ is the relative entropy of coherence \cite{Baumgratz,Streltsov}
\begin{equation}\label{coherence1}
C(\rho_{S})= S(\Delta_{H_{S}}(\rho_{S}) )-S(\rho_{S}),
\end{equation}
where
 $$\Delta_{H_{S}}(\rho_{S})=\sum_{i}\langle E_{i}| \rho_{S}| E_{i}\rangle | E_{i}\rangle\langle E_{i}|,$$
is the dephasing map, acting on the density matrix $\rho_{S}$, which removes all coherences from the various energy eigenspaces of $H_{S}$. According to Eq.\;(\ref{freedivided}), the entropy production is divided into two classical and quantum contributions:
\begin{equation}
\bar{\Delta S^{irr}}=\bar{\Delta S^{irr}_{C}}+\bar{\Delta S^{irr}_{Q}},
\end{equation}
where
\begin{equation}
\bar{\Delta S^{irr}_{C}}=S(p_{S}\Vert p^{eq}_{S})-S(p^{\prime}_{S}\Vert p^{eq}_{S}),
\end{equation}
\begin{equation}
\bar{\Delta S^{irr}_{Q}}=C(\rho_{S})-C(\rho^{\prime}_{S}).
\end{equation}
Now we aim to repeat the above process for the new expression of entropy production Eq.\;(\ref{epe2}) which obtained based on the new definition of heat. In this regard, the new entropy production will be divided to the classical and quantum parts as follows
\begin{equation}\label{epec}
\Delta S^{irr}_{C}=S(p_{S}\Vert p^{eq}_{S})-S(p^{\prime}_{S}\Vert p^{eq}_{S})-S(p^{\prime}_{B}\Vert p^{eq}_{B}),
\end{equation}
\begin{equation}\label{epeq}
\Delta S^{irr}_{Q}=C(\rho_{S})-C(\rho^{\prime}_{S})-C(\rho^{\prime}_{B}).
\end{equation}
To prove that these two contributions are both non-negative, we use the following equation:
\begin{eqnarray}
&&\Delta S^{irr}=I(\rho^{\prime}_{SB})\\\nonumber
& &=[I(\Delta_{H_{S}+H_{B}}(\rho^{\prime}_{SB})]+[I(\rho^{\prime}_{SB})-I(\Delta_{H_{S}+H_{B}}(\rho^{\prime}_{SB}))],
\end{eqnarray}
so, we have
\begin{equation}
\Delta S^{irr}_{C}=I(\Delta_{H_{S}+H_{B}}(\rho^{\prime}_{SB})),
\end{equation}
\begin{equation}
\Delta S^{irr}_{Q}=I(\rho^{\prime}_{SB})-I(\Delta_{H_{S}+H_{B}}(\rho^{\prime}_{SB})).
\end{equation}
Since these above equations are non-negative, therefore it could be concluded that Eq.\;(\ref{epec}) and Eq.\;(\ref{epeq}) are non-negative as well.\\
%%%%%%%%%%%%%%%%%%%%%%%%%%%%%%%%%%%%
\begin{figure}[t]
\centerline{\includegraphics[width=8cm]{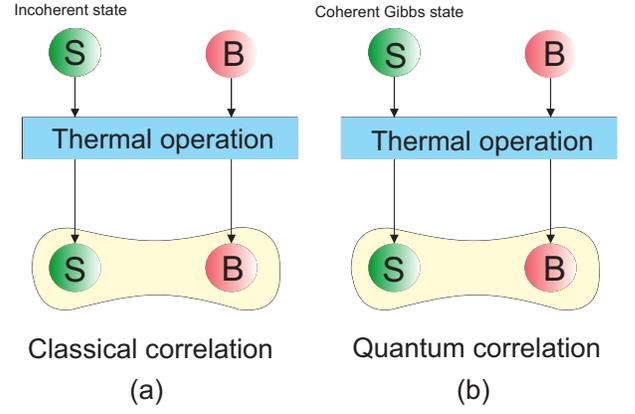}}
\vspace*{8pt}
\caption{(Color online). (a)  TO cannot generate quantum correlation from incoherent input states. (b) Conversely, if the input state of the system be coherent Gibbs state then correlation between the system and bath is quantum correlation.}\label{entropic}
\end{figure}
%%%%%%%%%%%%%%%%%%%%%%%%%%%%%%%%%%%%

A quantum operation is coherence preserving if and only if it is unitary and incoherent \cite{Peng}. With attention to this statement, the unitary operation $U$, which was mentioned in the definition of TO, is a coherence preserving operation. Thus, the total coherency of the both system and bath remains unchange under operate of $U$, i.e.,
\begin{equation}\label{coherence2}
C(\rho^{\prime}_{SB})=C(\rho_{S}\otimes\gamma_{B})=C(\rho_{S}).
\end{equation}
Substituting Eq.\;(\ref{coherence2}) into Eq.\;(\ref{epeq}), we have
\begin{equation}
\Delta S^{irr}_{Q}=C(\rho^{\prime}_{SB})-C(\rho^{\prime}_{S})-C(\rho^{\prime}_{B})=C_{cc}(\rho^{\prime}_{SB}),
\end{equation}
where $C_{cc}(\rho^{\prime}_{SB})$ is correlated coherence \cite{Kraft,Tan}. In contrast to \cite{Santos} where coherence of the bath contributes in entropy production, we see here that by using the new definition of heat only correlated coherence contributes and coherency of subsystems play no role. \\
Here it is interesting to note that if the initial state of the system is diagonal in energy bases, that is $C(\rho_{S})=0$, then
\begin{equation}\label{coherence3}
C(\rho^{\prime}_{S})=C(\rho^{\prime}_{B})=C_{cc}(\rho^{\prime}_{SB})=0,
\end{equation}
hence, the system and the bath will remain diagonal under the evolution. Also
\begin{equation}\label{ggh}
\Delta S^{irr}_{Q}=C_{cc}(\rho^{\prime}_{SB})=I(\rho^{\prime}_{SB})-I(\Delta_{H_{S}+H_{B}}(\rho^{\prime}_{SB}))=0,
\end{equation}
therefore
\begin{equation}
\Delta S^{irr}=\Delta S^{irr}_{C}=I(\Delta_{H_{S}+H_{B}}(\rho^{\prime}_{SB})),
\end{equation}
this means that the entropy production has only a classical contribution.
In other words, we can say that if the initial state of the system is diagonal in energy basis, then it is impossible to create a quantum correlation between the system and bath by using TO. Another interesting feature is that if the initial state of the system is in the coherent Gibbs state, which is defined as follows \cite{Rudolph,Kwon}
\begin{equation}\label{coherence gibs state}
\vert \lambda\rangle:=\sum_{i}\sqrt{\frac{e^{-\beta E_{i}}}{Z_{S}}}\vert E_{i}\rangle ,
\end{equation}
from Eq.\;(\ref{epec}) one has $\Delta S^{irr}_{C}= 0$. Therefore, only quantum correlation contributes in entropy production
\begin{equation}\label{ss}
\Delta S^{irr}=\Delta S^{irr}_{Q}=C_{cc}(\rho^{\prime}_{SB}).
\end{equation}
Consequently, if the initial state of the system is a coherent Gibbs state, then under TO if there is a correlation between the system and the bath, this correlation must be quantum correlation, which is in complete contrast with the previous case.

\section{Conclusion}\label{conclusion}\label{Sec5}
In this work, it was shown that changing the definition of heat results in the modified version of the entropy production. Specifically for our case, applying the new definition of heat, one could found an extra modified term for the entropy production with respect to the familiar form of the entropy production. It was determined that the effect of this modified term is eliminated for large environments, then both entropy productions come close to each other.  \\
In addition it was demonstrated that using the new definition of heat, if one assumes that the initial state of the system and the bath is product and also the map is entropy preserving then the entropy production is equal to mutual information between the system and the bath. \\
There is another feature which is resulted from utilizing the new definition of heat is that the boundary of free energy variation under TO could affectively be changed, so that with respect to the usual case, the boundary of the free energy under TO becomes tighter.\\
Also, the role of  quantum coherence in entropy production was studied, it was shown that if the initial state of the system is in the diagonal energy bases, then using the TO we cannot create a quantum correlation between the system and the bath. While, if the initial state of the system is coherent Gibbs state, the correlation between system and bath must be quantum correlation.

\appendix

\section{Obtain the Equations}
The following relations are true for thermal operations:
\begin{equation}
\Delta E_{S}+\Delta E_{B}=0,
\end{equation}
\begin{equation}
I(\rho^{\prime}_{SB})=\Delta S_{S}+\Delta S_{B},
\end{equation}
\begin{equation}
\Delta F_{B}=TS(\rho^{\prime}_{B}\Vert\rho^{eq}_{B}),
\end{equation}
where the first is because of the energy conservation condition, the second relation is due to the fact that the initial total state is a direct product of the system and the bath and the third relation comes from this condition that the initial state of the bath is thermal. \\
In the following lines, more detail about the equations of Sec. III is presented so that the approach is slightly different from previous methods \cite{Deffner,Deffner2,Esposito,Esposito2,chen}.

Derivation of Eq.\;(\ref{epe}):
\begin{eqnarray}
\bar{\Delta S^{irr}}&=&\Delta S_{s}-\beta\Delta \bar{Q}=\Delta S_{s}+\beta\Delta E_{B}\\\nonumber
&  & =\Delta S_{s}-\beta\Delta E_{S}=-\beta\Delta F_{S},
\end{eqnarray}
here we use $\Delta \bar{Q}=-\Delta E_{B}$ for the definition of heat.

Derivation of Eq.\;(\ref{epem}):
\begin{eqnarray}
\bar{\Delta S^{irr}} &=&-\beta\Delta F_{S}=\Delta S_{s}-\beta\Delta E_{S}\\\nonumber
& &=I(\rho^{\prime}_{SB})-\Delta S_{B}+\beta\Delta E_{B}\\\nonumber
& &=I(\rho^{\prime}_{SB})+\beta\Delta F_{B}=I(\rho^{\prime}_{SB})+S(\rho^{\prime}_{B}\Vert\rho^{eq}_{B}).
\end{eqnarray}

Derivation of Eq.\;(\ref{epe2}):
\begin{eqnarray}
\Delta S^{irr}&=&\Delta S_{s}-\beta\Delta Q=\Delta S_{s}+\beta(\Delta E_{B}-\Delta F_{B})\\
&&=\Delta S_{s}-\beta\Delta E_{S}-\beta\Delta F_{B}=-\beta(\Delta F_{S}+\Delta F_{B}),\nonumber
\end{eqnarray}
where we use $\Delta Q=\frac{-1}{\beta} \Delta S_{B}$ for the definition of heat.

Derivation of Eq.\;(\ref{epem2}):
\begin{eqnarray}
\Delta S^{irr}&=&-\beta(\Delta F_{S}+\Delta F_{B})\\
&&=\Delta S_{s}+\Delta S_{B}-\beta(\Delta E_{S}+\Delta E_{B})=I(\rho^{\prime}_{SB}).\nonumber
\end{eqnarray}

 %=============================================================%
%=============================================================%
%=======================  References =========================%
%=============================================================%
%=============================================================%

%%The Appendices part is started with the command \appendix;
%%\appendix


%% References


%\bibliography{sample.bib}



%%%%%%%%%%%%%%%%%%%%%%%%%%%%%%%%%%%%%%%%%%%%%%%%%%%%%%%%%%%%%%%%%%%%%
%%%%%%%%%%%%%%%%%%%%%%%%%%%%%%%%%%%%%%%%%%%%%%%%%%%%%%%%%%%%%%%%%%%%

\end{document}